\documentclass[aps,preprint]{revtex4-1}
\usepackage{hyperref}
\hypersetup{colorlinks=true, urlcolor=blue}

\usepackage{epsfig}
\usepackage{newlfont}
\usepackage{amssymb}
\usepackage{amsfonts}
\usepackage{amsmath}
\usepackage{bm}

\newcommand{\Tr}{\mbox{Tr}}
\newcommand{\ket}[1]{|{#1}\rangle}
\newcommand{\bra}[1]{\langle{#1}|}

\def\lsim{\mathrel{\rlap{\lower4pt\hbox{\hskip1pt$\sim$}}
    \raise1pt\hbox{$<$}}}                
\def\gsim{\mathrel{\rlap{\lower4pt\hbox{\hskip1pt$\sim$}}
    \raise1pt\hbox{$>$}}}                

\begin{document}

\renewcommand*{\DefineNamedColor}[4]{%
   \textcolor[named]{#2}{\rule{7mm}{7mm}}\quad
  \texttt{#2}\strut\\}

\definecolor{red}{rgb}{1,0,0}
\definecolor{cyan}{cmyk}{1,0,0,0}

\title{Transmitting quantum information by superposing causal order\\of mutually unbiased measurements}

\author{Manish K. Gupta}
\email{manishh.gupta@gmail.com}

\author{Ujjwal Sen}
\email{ujjwal@hri.res.in}

\affiliation{Quantum Information and Computation Group,
Harish-Chandra Research Institute, Chhatnag Road, Jhunsi, Prayagraj (Allahabad) 211 019, India
}


\begin{abstract}
Two quantum measurements sequentially acting one after the other, if they are mutually unbiased, will lead to a complete removal of information encoded in the input quantum state. We find that if the order of the two sequential measurements can be superposed, with a quantum switch, then the information encoded in the input can still be retained in the output state. 
\end{abstract}


\maketitle 


\noindent\emph{Introduction.--} 
Quantum mechanics, in its usual considerations, implicitly assumes that space and time stand fixed, and the events take place with a definite causal order. If we forgo the idea of the definite ordering of events and assume the validity of quantum mechanics locally, then quantum mechanics allows quantum particles to experience coherent superposition of alternative evolutions \cite{Aharonov90,Oi03}, or to experience a set of evolutions in the superposition of alternative orders \cite{Chiribella13,Chiribella12}. In principle, quantum mechanics does allow for such superpositions. For example, if a particle travels through a superposition of alternative paths, the interference of noisy processes on different paths leads to a cleaner communication channel \cite{Gisin05,Abbott18}. Similarly, it has been observed that when a particle experiences a noisy process in the superposition of orders, the interference between the alternative ordering of processes leads to boost in the capacity to communicate classical and quantum bits \cite{Ebler18,Salek18,ChiribellaBanik18,Chiribella19}. Chiribella \emph{et al.} \cite{Chiribella13} introduced the notion of \emph{quantum switch}, which allows  \emph{two different orderings} of processes to be superposed, giving rise to a feature called causal non-separability \cite{brukner12,Mateus15,Oreshkov16}. In recent years, the application of quantum switch and other causally non-separable processes have led to quantum advantage in various tasks, such as testing properties of quantum channels \cite{Chiribella12,Mateus14}, winning nonlocal games \cite{brukner12}, metrology \cite{ChiranjibManish18}, improving teleportation protocols in noisy scenarios \cite{ChiranjibPati19}, and reducing quantum communication complexity \cite{Guerin16}. The advantage of a quantum switch has been experimentally demonstrated in various photonic setups \cite{Walther15,Walther17,White18}.

In this paper, we investigate the effect of superposing two alternative orderings of non-commuting sequential measurements on a quantum system. Measurement in quantum mechanics connects the quantum world to the classical one. Any measurable property, known as observable of a quantum system, is attached to a measurement operator, say $\hat{M}$, and a definite value of an observable can only be assigned to a system when the system is an eigenstate of the operator $\hat{M}$. Two observables, $M$ and $N$, of a system can simultaneously be assigned values only when the system is in an eigenstate of both the operators $\hat{M}$ and $\hat{N}$. In this work, we show that when the two different causal orderings of two non-commuting mutually unbiased measurement operators are allowed to act on a quantum system coherently, using a quantum switch, then the resulting state still has quantum information about the input. This is strikingly different from the situation when two orders of the sequential measurements act incoherently on the input state, and in such cases, the output does not have any information about the input quantum state. We first analyze the set-up where the Hilbert space dimension is two, and later generalize the considerations to higher dimensions. 

\begin{figure}
\begin{center}
\epsfig{figure = 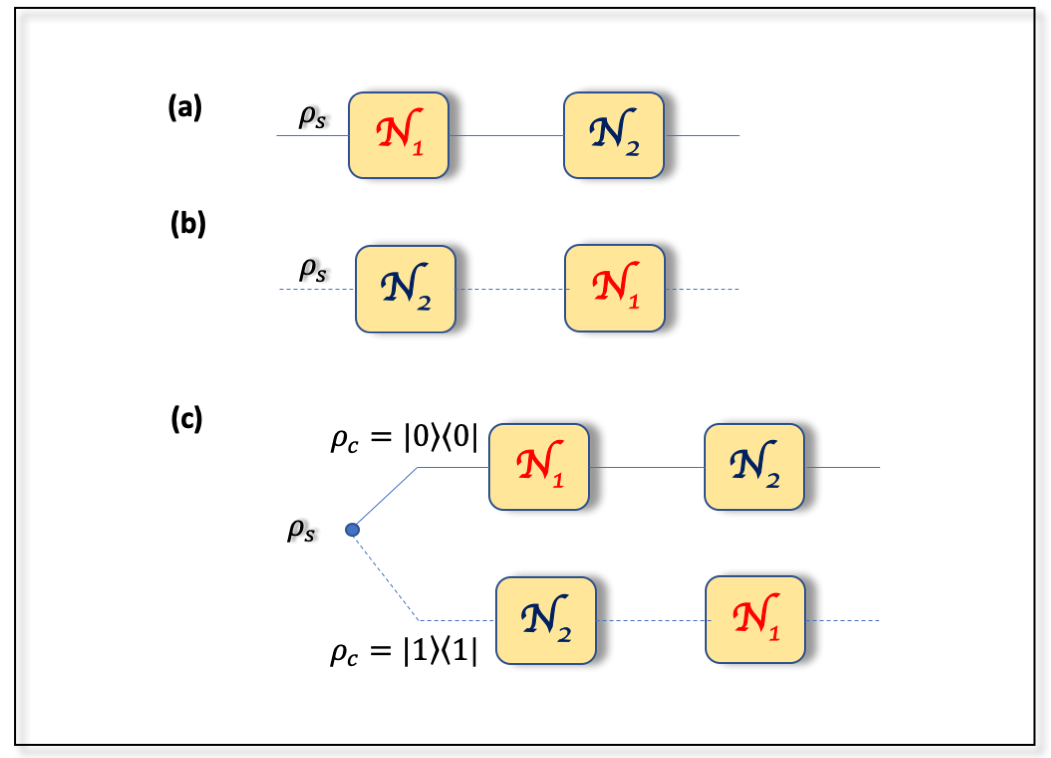, height=.3\textheight, width=0.6\textwidth,angle=-0}
\caption{Superposing two different ordering of processes. a) The first ordering is where $\mathcal{N}_{1}$ and then $\mathcal{N}_{2}$ acts on an input quantum state $\rho_{s}$. b) The other ordering is where the process $\mathcal{N}_{2}$ and then $\mathcal{N}_{1}$ acts on the input quantum state $\rho_{s}$. c) The two different orderings of the processes, superposed using a control qubit $\rho_{c}$, act on the input quantum state $\rho_{s}$.}
\label{fig:1}
\end{center}
\end{figure}

\noindent\emph{Fourier-inverted measurements and a quantum switch.--} Consider a Hilbert space $\mathcal{H}$ with $\dim \mathcal{H} = d$, and two sets of mutually unbiased bases (MUBs) \cite{Wootters89,Somshubhro02,Gilad14} of it. Two sets of mutually orthonormal vectors, $\left\lbrace\ket{e_{i}}\right\rbrace^{d}_{i=1}$ and $\lbrace\ket{f_{j}}\rbrace^{d}_{j=1}$, are said to be MUBs if $| \bra{e_{i}}\ket{f_{j}}|=\frac{1}{\sqrt{d}}, \forall~i,j=1,\dotsc,d$. This can be achieved in any dimension by considering $\left\lbrace \ket{f_{j}} \right\rbrace$ to be Fourier-inverted vectors of the set $\left\lbrace \ket{e_{j}} \right\rbrace$. Using the MUBs, we define positive operator valued measure (POVM) operators, $\left\lbrace E_{i} \right\rbrace$, $i=1,\dotsc,d$, and $\left\lbrace F_{j} \right\rbrace$, $j=1,\dotsc,d$, satisfying $\sum^{d}_{i=1} E_{i}=\mathcal{I}$ and $\sum^{d}_{j=1} F_{j}=\mathcal{I}$. Specifically, $E_{i}=\ket{e_{i}}\bra{e_{i}}$ and $F_{j}=\ket{f_{j}}\bra{f_{j}}$ $\forall~i,j=1,\dotsc,d$. Here, $\mathcal{I}$ denotes the identity operator on the Hilbert space $ \mathcal{H}$. An arbitrary element of $\left\lbrace E_{i} \right\rbrace$ does not commute with the same of $\left\lbrace F_{i} \right\rbrace$. These POVMs actually belong to the restricted class of projective measurements. The two POVMs can act sequentially on a state $\rho_{s}$ on $\mathcal{H}$. The two sets of sequential measurements have two physically different orderings. 

A quantum switch superposes the two orderings of the measurement processes, by using a control qubit, as schematically depicted in Fig.~\ref{fig:1}. If the two processes $\mathcal{N}_{1}$ and $\mathcal{N}_{2}$ are completely depolarizing channels, then a coherent superposition of the channels still allows some information to be passed from input to output \cite{Ebler18,Salek18,ChiribellaBanik18}. Intuitively, it is the non-commutativity of the Kraus elements of the maps $\mathcal{N}_{1}$ and $\mathcal{N}_{2}$ that leads to the ``super-activation". We ask the question whether non-commutativity of measurement elements can also lead to such a super-activation?

The two orders of the non-commuting sequential measurement elements, $E_{i}F_{j}$ and $F_{j}E_{i}$, can be applied coherently on a state $\rho_{s}$ using a control qubit $\rho_{c}$. The Kraus operators for the process are given as
\begin{align}
\mathcal{W}_{ij}=\ket{0}\bra{0} \otimes  E_{i}F_{j}  +  \ket{1}\bra{1} \otimes F_{j}E_{i} , \\
\sum_{i,j} \mathcal{W}_{ij}^\dagger \mathcal{W}_{ij}= \mathcal{I}_{4},
\end{align} 
with $\mathcal{I}_{4}$ denoting the identity operator on $\mathbb{C}^4$. $\mathcal{W}_{ij}$ represents the measurement outcomes when the two non-commuting measurement elements, $E_{i}$ and $F_{j}$, act sequentially on a quantum state $\rho_{s}$ in two different orders, coherently controlled by using a control qubit $\rho_{c}$. The final state after coherent application of the two sequential POVMs is%
\begin{align}
\mathcal{N} \left( \rho_{c} \otimes \rho_{s}\right) = \sum_{i,j}\mathcal{W}_{ij} \left( \rho_{c} \otimes \rho_{s} \right) \mathcal{W}_{ij}^\dagger.
\end{align}


\noindent\emph{Two-dimensional Fourier-inverse measurements.--} Let us consider the two-dimensional complex Hilbert space, in which $E = \{   \ket{0},\ket{1}\}$ and $F = \{ \ket{+}, \ket{-}\}$ are two MUBs. The corresponding unbiased measurement operators are $E_{1}=\ket{0}\bra{0}$, $E_{2}=\ket{1}\bra{1}$ and $F_{1}=\ket{+}\bra{+}$, $F_{2}=\ket{-}\bra{-}$. Here, $\ket{0}$ and $\ket{1}$ are eigenvectors of the Pauli $\sigma_{z}$ matrix, while $\ket{+}$ and $\ket{-}$ are the same of the Pauli $\sigma_{x}$. The two vector sets corresponding to the non-commuting measurement operators are Fourier inverses of each other. Consider now an arbitrary state $\ket{\psi}= \alpha\ket{0}+\beta\ket{1}$, of this system, measured in the $\left\lbrace\ket{0},\ket{1}\right\rbrace$ basis. The output will be the state $\ket{0}\bra{0}$ with probability $|{\alpha}|^2$ and the state $\ket{1}\bra{1}$ with probability $|{\beta}|^2$. If subsequently measured in the $\left\lbrace \ket{+},\ket{-}\right\rbrace$ basis, then the output state would be the completely mixed state. Similarly, if the state $\ket{\psi}$ is first measured in the $\left\lbrace \ket{+},\ket{-}\right\rbrace$ basis, the output will be the state $\ket{+}\bra{+}$ with probability $\left( \alpha + \beta \right)^2/4$ and the state $\ket{-}\bra{-}$ with probability $\left( \alpha - \beta \right)^2/4$. Subsequently, if the resultant state is measured in the $\left\lbrace\ket{0},\ket{1}\right\rbrace$ basis, then the output state would again be the completely mixed state. Consecutive complementary measurements erase all information in the input quantum state. This for example is potentially related to the security of the Bennett Brassard 1984 quantum cryptography protocol \cite{BB84} (see also \cite{E91,B92,Gisin02}), where preparation in different MUBs are used to defeat the eavesdropper. However, as we will find out, a superposition of these two measurement processes leads to an output that does contain non-trivial quantum information about the input state.

A general qubit state on the Bloch sphere can be represented as $\rho_{s} = \frac{1}{2} (\mathcal{I}_{2} + \overrightarrow{r} \cdot  \overrightarrow{\sigma})$, where $\mathcal{I}_{2}$ is the identity operator on $\mathbb{C}^2$, $\overrightarrow{\sigma}$ is the three-element vector of Pauli matrices $ (\sigma_{x},\sigma_{y},\sigma_{z})$, and $\overrightarrow{r} = (r_{x},r_{y},r_{z})=(\cos{\phi}\sin{\theta},\sin{\phi}\sin{\theta},\cos{\theta})$ is the unit Bloch vector. The control qubit is fixed to the state $\rho_{c} = \ket{\psi_{c}}\bra{\psi_{c}}$, where $\ket{\psi_{c}} = \sqrt{p} \ket{0} + \sqrt{1-p} \ket{1}$ and $p$ is a parameter that regulates the coherence of the control state. 

When the two measurements operate in a superposition of different orders, $E_{i}F_{j}$ and $F_{j}E_{i}$, on a state $\rho_{s}$, using a quantum switch, the output state is given as
\begin{align}
\mathcal{N} \left( \rho_{c} \otimes \rho_{s}\right) &= \sum_{i,j}\mathcal{W}_{ij} \left( \rho_{c} \otimes \rho_{s} \right) \mathcal{W}_{ij}^\dagger \nonumber \\
&= \sum_{i,j}\mathcal{W}_{ij} \left(  \rho_{c} \otimes \frac{1}{2} (\mathcal{I}_{2} + \overrightarrow{r} \cdot  \overrightarrow{\sigma}) \right) \mathcal{W}_{ij}^\dagger \nonumber \\
&= \frac{1}{2} \left( \rho_{c} \otimes \mathcal{I}_{2}   + \sum_{i,j} \mathcal{W}_{ij} \left(  \rho_{c} \otimes  \overrightarrow{r} \cdot  \overrightarrow{\sigma}  \right) \mathcal{W}_{ij}^\dagger  \right).
\end{align}
After simplifying, we get the output density matrix as%
\begin{align}
\label{eq:outputRho}
&\mathcal{N} \left( \rho_{c} \otimes \rho_{s}\right) = \nonumber \\
&\left(
\begin{array}{cccc}
 \frac{p}{2} & 0 & \alpha (\cos{\theta} +1) & \alpha \sin{\theta} e^{\iota\phi} \\
 0 & \frac{p}{2} & \alpha \sin{\theta} e^{-\iota\phi} & - \alpha (\cos{\theta}-1) \\
 \alpha (\cos{\theta}+1) & \alpha \sin{\theta} e^{\iota\phi} & \frac{1-p}{2} & 0 \\
 \alpha \sin{\theta} e^{-\iota\phi} & -\alpha (\cos{\theta}-1) & 0 & \frac{1-p}{2} \\
\end{array}
\right),
\end{align}%
where $\alpha=\frac{1}{4}\sqrt{(1-p) p}$. The distance, $D\left( \Tr_{c}~\mathcal{N}\left( \rho_{c} \otimes \rho_{s}\right)~||~ \frac{1}{2} \mathcal{I}_{2} \right)$, vanishes, as the state after tracing out the control qubit from $\mathcal{N}\left( \rho_{c} \otimes \rho_{s}\right)$ is a completely mixed state. Although $ \Tr_{c}~\mathcal{N}\left( \rho_{c} \otimes \rho_{s} \right)$ does not have any information about the input state, $\mathcal{N}\left( \rho_{c} \otimes \rho_{s} \right)$ does have. And the amount of that information can be quantified by $D\left(\mathcal{N}\left( \rho_{c} \otimes \rho_{s}~||~ \rho_{c} \otimes \frac{1}{2} \mathcal{I}_{2} \right) \right)$, as $\rho_{c} \otimes \frac{1}{2} \mathcal{I}_{2}$ represents the joint state when the effect of the control qubit is absent.

The output density matrix after the coherent sequential measurement has off-diagonal terms, in contrast to the case of incoherent sequential measurements. The control qubit is made incoherent by setting $p=1$ or $p=0$, which in turn makes the off-diagonal terms of the output density matrix in Eq.~(\ref{eq:outputRho}) to vanish. This means, in particular, that when the two orders of sequential measurements act incoherently on the input state, then the final state after the measurement will have only diagonal terms in the density matrix. The same happens when we choose $\rho_{c} =p \ket{0}\bra{0} + (1-p) \ket{1}\bra{1}$, where $0 \leq p \leq 1$.

\noindent\emph{Fourier-inverted measurements for qudits.--} We now consider a $d$-dimensional Hilbert space with MUBs, $\left\lbrace\ket{\phi^{b}_{n}} \right\rbrace$, where $b$ labels the basis and $n=1,2,\dots,d$ labels the vectors within a basis. The number of values that $b$ can assume is not known, but for every dimension, $d$, one can always choose a computational basis and its Fourier inverse, which will be MUBs. The set of projectors that can be defined using the $b^{\text{\tiny{th}}}$ basis vectors, $\left\lbrace M^{b}_{n}=\ket{\phi^{b}_{n}}\bra{\phi^{b}_{n}} \mid n=1,2,\dots,d\right\rbrace$, form a set of measurement elements, $M^{b}_{n} \geq 0, \forall n,$ and $\sum_{n}M^{b}_{n}=1, \forall~b,$.

Now, using any pair of measurement operator $M^{k}_{i}$ and $M^{l}_{j}$, defined using the $k^{\text{\tiny{th}}}$ and $l^{\text{\tiny{th}}}$ MUBs, we can construct two causal orders of sequential measurements, viz. $\left\lbrace M^{k}_{i}M^{l}_{j} \right\rbrace_{ij}$ and $\left\lbrace M^{l}_{j}M^{k}_{i}\right\rbrace_{ji}$. The two orders of the non-commuting sequential measurements can be applied coherently on a state $\tilde{\rho}_{s}$ using the quantum switch discussed earlier. The Kraus operators for the process are given by
\begin{align}
\widetilde{\mathcal{W}}_{ij}=\ket{0}\bra{0} \otimes M^{k}_{i}M^{l}_{j}   +  \ket{1}\bra{1} \otimes M^{l}_{j}M^{k}_{i}, \\
\sum_{i,j} \widetilde{\mathcal{W}}_{ij}^\dagger \widetilde{\mathcal{W}}_{ij}= \mathcal{I}_{2d},
\end{align} 
with $\mathcal{I}_{2d}$ being the identity operator on $\mathbb{C}^{2d}$. The  state after the action of the quantum switch is given as
\begin{align}
\widetilde{\mathcal{N}} \left( \rho_{c} \otimes \tilde{\rho}_{s} \right) = \sum_{i,j}\widetilde{\mathcal{W}}_{ij} \left( \rho_{c} \otimes \tilde{\rho}_{s} \right) \widetilde{\mathcal{W}}_{ij}^\dagger,
\end{align}
where $\tilde{\rho}_{s}$ is a general quantum state on $\mathbb{C}^{d}$.

It can be shown, via explicit analytical calculations, for Hilbert spaces of relatively small dimensions that the state after application of the quantum switch is not diagonal and the final state depends on the coherence of the control qubit. Therefore, if the two non-commuting sequential measurements are applied coherently on the input state, then the output density matrix will have off-diagonal elements, containing quantum information of the input. Otherwise, the density matrix will be the identity matrix, as seen in case of the two-dimensional Hilbert space.

\noindent\emph{Discussion.--} We consider a setup where two non-commuting measurement operators are made to act on a quantum state in two different orders coherently. We find that when the two sequential orderings of measurement operators are applied coherently on a quantum state using a quantum switch, then the output state has non-zero off-diagonal elements. This is surprising when compared to the case where the two orderings of measurement operators are applied incoherently, as for the latter, the off-diagonal elements are zero.

We know from the conceptualization of MUBs that when a system in any state is measured in one basis, the probability distribution upon measuring it in a second measurement basis, with the two bases being mutually unbiased, is completely random \cite{Wootters89,Somshubhro02,Gilad14}. It means, in particular, that the final density matrix of the system will be diagonal. We find that when the sequential measurement order is coherently superposed using a quantum switch, then the output density matrix has non-zero off-diagonal elements.

This result would potentially have an impact on quantum tasks such as the quantum key distribution \cite{BB84, E91,B92,Gisin02} where preparation of the encoding states in different MUBs is used to baffle the eavesdropper.
Although we have used mutually unbiased measurements, which are projective measurements, the set-up can potentially be extended to any pair of non-commuting generalized measurements.


\end{document}